\documentclass[prd,onecolumn,nofootinbib,preprintnumbers,floatfix]{revtex4}
\usepackage[plainpages=false, colorlinks=true, anchorcolor=blue, linkcolor=blue, citecolor=blue, bookmarks=false]{hyperref}
\usepackage[utf8]{inputenc}
\usepackage{amsfonts,amsmath,amssymb}
\usepackage{graphicx}
\usepackage{natbib}
\usepackage{enumitem}
\usepackage{subcaption}
\usepackage{comment}
\usepackage{caption}
\usepackage{lipsum}
\newcommand{\rthis}[1]{\textcolor{black}{#1}}

\begin{document}
\title{Comparison of $R_h=ct$ and $\Lambda$CDM using DESI DR1 measurements}
\author{Kunj \surname{Panchal}}\altaffiliation{E-mail: ep21btech11017@iith.ac.in}

\author{Shantanu  \surname{Desai}}  
\altaffiliation{E-mail: shntn05@gmail.com}

\begin{abstract}
We use  DESI DR1 BAO measurements  of the ratio of tranvserse comoving distance to Hubble distance in order to test the compatibility  of $R_h=ct$ over flat  $\Lambda$CDM. For this purpose, we used Bayesian model selection to evaluate the efficacy of these models given the observed data.
When we consider the BAO measurements up to redshift of  1.3, both models are equally favored. However, when we consider the Lyman-$\alpha$ QSO measurement at redshift of 2.33, we find Bayes factors of greater than  100 for flat $\Lambda$CDM over $R_h=ct$ using two different priors for $\Omega_m$, indicating that $\Lambda$CDM is decisively favored over $R_h=ct$. The same is the case when we combine all the measurements. Therefore, the DESI DRI  measurements rule out $R_h=ct$ cosmology, albeit this is driven  by   the Lyman-$\alpha$ QSO measurement at $z=2.33$.

\end{abstract}  
 \affiliation{Department of Physics, Indian Institute of Technology, Hyderabad, Telangana-502284, India}
 
 \maketitle
\section{Introduction}
The current concordance $\Lambda$CDM model of cosmology consisting of  70\% dark energy and 25\% cold (non-baryonic) dark matter and 5\% baryons~\cite{Ratra03} agrees with  CMB and LSS  at large scales~\cite{Planck2018}. There are however numerous data-driven tensions which have come up in recent years such as the Hubble constant tension~\cite{Lahav,DiValentino,Verde,Bethapudi,Smoot}, $\sigma_8$ tension~\cite{Abdalla22}, failure to detect CDM candidates in laboratory based experiments~\cite{Merritt}, Lithium-7 problem in Big-Bang nucleosynthesis~\cite{Fields}, detection of very massive galaxies at high redshifts~\cite{Boylan}, etc. An up-to-date status report of all problems and anomalies in $\Lambda$CDM  can be found in ~\cite{Periv,Abdalla22,Peebles22,Banik}.

One of the alternatives  to the $\Lambda$ CDM model is the $R_h=ct$ universe model, proposed by Melia~\cite{Melia07,Shevchuk,Melia2012}. In this model, the size of the Hubble sphere given by $R_h(t)=ct$ is the same at all  times.  This model has the Hubble parameter ($H(z)$) given by  $H(z)=H_0(1+z)$, and the cosmic scale factor  $a(t) \propto t $. A large number of  works by Melia and collaborators have found that this model is in better agreement than $\Lambda$CDM using a plethora of observations, such as cosmic chronometers~\cite{Meliamaier,Meliachrono}, quasar core angular size measurements~\cite{Meliaquasar}, quasar X-ray and UV fluxes~\cite{MeliaUV}, galaxy cluster gas mass fraction measurements~\cite{Meliafgas}, Type Ia SN~\cite{MeliaSN}, strong lensing~\cite{MeliaSL,Melialensing23}, FRBs~\cite{MeliaFRB}, JWST observations of highest redshift quasars and Einstein ring~\cite{MeliaJWST,MeliaJWST1}. However, other researchers have reached opposite conclusions by analyzing multiple observational probes~\cite{Shafer,Seikel_2012_rhct,LewisBBN,Haridasu1,Lin,Hu2018,Tu2019,Fuji,Haveesh,Panchalfgas}.

One criticism of some of the above works disputing $R_h=ct$, which drew their conclusions based on Baryon Acoustic Oscillation (BAO) and Alcock-Paczynski effects~\cite{Seikel_2012_rhct,Haridasu1} is that the BAO measurements were scaled by the size of the sound horizon at the drag epoch ($r_d$), which implicitly assumes the $\Lambda$CDM model~\cite{Meliamaier,Meliachrono}. \rthis{The first model-independent test of $R_h=ct$ and other power-law cosmologies using BAO measurements  was carried out in ~\cite{Shafer}, which assumed $r_d$ to be a free parameter and used only the relative distances from BAO measurements.}
This was followed up in a series of papers by Melia~\cite{MeliaBAO17,MeliaBAO20,MeliaBAO22}  with the latest result in ~\cite{Melia23} (M23 hereafter).
M23 considered the ratios of transverse comoving distance ($D_m$) and Hubble distance ($D_H$) using the eBOSS Lyman-$\alpha$ measurements at $z=2.334$, so that there is no dependence on $H_0$ and $r_d$.  
M23 considered four different cosmological models: $R_h=ct$, Milne Universe, Einstein-de Sitter universe, flat $\Lambda$CDM using Planck cosmology. They found that the Milne and Einstein-DeSitter universe are strongly ruled out by the data. M23 also deduced that $R_h=ct$ is favored compared to Planck $\Lambda$CDM with $p$-values of 0.39 and 0.03 respectively. This was also confirmed using Bayesian model comparison.

In this work, we implement the same tests as in M23 using the DESI 2024 BAO data release (also known as DESI DR1)\footnote{At the time of writing, the DESI collaboration has recently published another data release, DR II~\cite{DESI2025}. However, for this manuscript it is sufficient to use DESI DRI results.}  This manuscript is structured as follows.  The DESI 2024 DR1  dataset used for the analysis is discussed in Sect.~\ref{sec:data}. A brief primer on Bayesian model comparison can be found in Sect.~\ref{sec:bayes}. Our analysis methodology and results are presented  in Sect.~\ref{sec:results}. We conclude in Sect.~\ref{sec:conclusions}.

\section{DESI 2024 dataset}
\label{sec:data}
The DESI collaboration did their first release (DR1) of the BAO data in April 2024~\cite{DESI2024} using around six million extragalactic objects in the redshift range: $0.1<z<4.2$. The target samples used consisted of  bright galaxy sample (BGS), the luminous red galaxy (LRG) sample, the emission line galaxy sample (ELG), Quasar sample (QSO), and Lyman-$\alpha$  (Ly$\alpha$) sample.  
The LRG sample was further subdivided into three redshift intervals
spanning the range: $0.4 < z < 0.6$, $0.6 < z < 0.8$, $0.8 < z < 1.1$, referred to as LRG1, LRG2, and LRG3, respectively. The ELG sample was also divided into two redshift ranges, $0.8<z<1.1$ and $1.1<z<1.6$, referred  to as ELG1 and ELG2, respectively. 
The DESI Collaboration provided  measurements of $D_m/r_d$ and $D_H/r_d$  for LRG1, LRG2, LRG3+ELG1, ELG2, and Lyman-$\alpha$ 
along with the  correlation coefficient ($r$). These values along with their effective redshifts can be found in Table I of ~\cite{DESI2024}.
For the remaining samples (BGS and QSO), only $D_V/r_D$ was provided due to the lower signal-to-noise ratio, where $D_V$ corresponds to the angle-averaged distances quantifying the average distances along and perpendicular to the line of sight. We do not use this measurement for our analysis, as one cannot do model-independent tests. 

\section{Brief primer on Bayesian Model Comparison}
\label{sec:bayes}
We provide  an abridged introduction   to Bayesian model comparison used to compare the relative efficacy of $\Lambda$CDM and $R_h=ct$. More details can be found  in recent reviews~\cite{Trotta,Weller,Sanjib,Krishak}.
To evaluate the significance of one  model ($M_2$) as compared to another one ($M_1$), we calculate the Bayes factor ($B_{21}$) given by:
\begin{equation}
B_{21}=    \frac{\int P(D|M_2, \theta_2)P(\theta_2|M_2) \, d\theta_2}{\int P(D|M_1, \theta_1)P(\theta_1|M_1) \, d\theta_1} ,  \label{eq:BF}
\end{equation}
where $P(D|M_2,\theta_2)$ is the likelihood for the model $M_2$ given the data $D$ and $P(\theta_2|M_2)$ denotes the prior on the parameter vector $\theta_2$ for the model $M_2$.  The denominator denotes the same for model $M_1$. The full expression in the numerator and denominator is referred to as Bayesian evidence for each of the two models. 
If $B_{21}$ is greater than one, then $M_2$ is  preferred over $M_1$ and vice-versa. The significance can be qualitatively assessed depending on the numerical value of the Bayes factor based on the Jeffreys' scale~\cite{Trotta}. 

\section{Analysis and Results}
\label{sec:results}
\begin{figure}
    \centering
    \includegraphics[width=\linewidth]{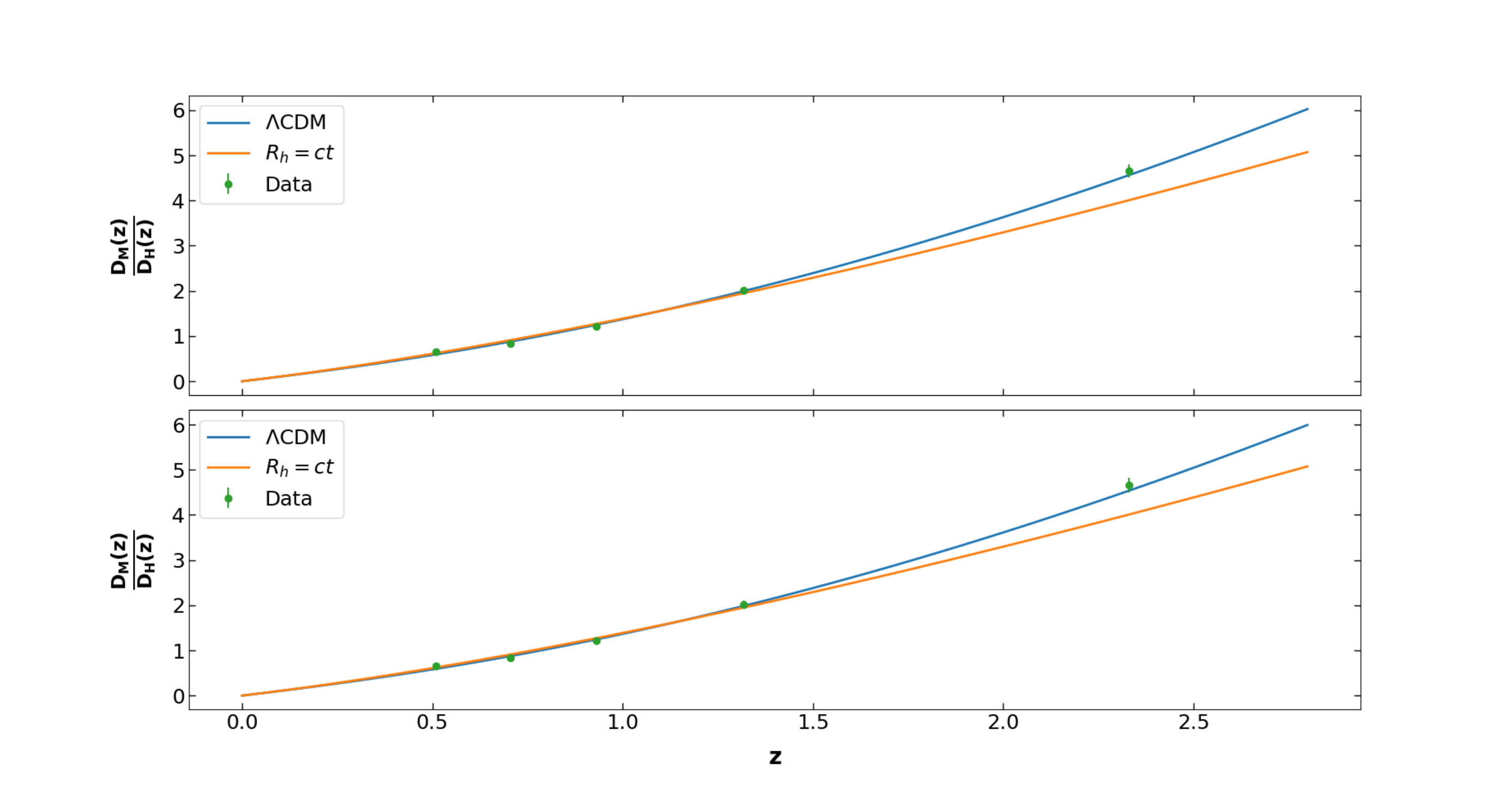}
    \caption{The best-fit plots for the  ratio $\frac{D_M(z)}{D_H(z)}$ plotted for $\Lambda$CDM and $R_h = ct$ cosmologies. The upper panel shows the plots for the case when correlation in the ratio of $D_M$ and $D_H$ is taken into account while the bottom  panel shows the same without including the correlation between the measurements. For $\Lambda$CDM, the plots show the best-fits using Normal priors, with the best fit value of $\Omega_m$ being 0.32 and 0.31, for correlation and no correlation case, respectively. $R_h = ct$ on the other hand has no free parameters.}
    \label{fig:ratioplot}
\end{figure}

We now write expressions for our observables $D_H(z)$ and $D_M(z)$ for both $\Lambda$CDM and $R_h=ct$. Both these quantities depend on the Hubble parameter $H(z)$, which is given by: 
\begin{eqnarray}
H(z) &=& H_0\sqrt{\Omega_M (1+z)^3+ 1-\Omega_M}~\text{ for flat $\Lambda$CDM } \\
\label{eq:lcdm}
H(z) &=& H_0 (1+z)~\text{ for $R_h=ct$} 
\label{eqrh=ct}
\end{eqnarray}

The expression for the Hubble distance is given by:
\begin{equation}
D_H(z) = \frac{c}{H(z)}
\label{eq:dhz}
\end{equation}
The expression for the transverse comoving distance is given by
\begin{equation}
D_M(z) = c \int_0^z \frac{1}{H(z)}
\label{eq:dmz}
\end{equation}
For our observables, we use the ratio of   the estimated values of  $D_m/r_d$ and $D_h/r_d$ from DESI DR1,
and then similar to M23 compare  to the ratio of Eq.~\ref{eq:dmz} and  Eq.~\ref{eq:dhz}. As discussed in M23, although the BAO measurements of $D_H$ and $D_M$ assume a fiducial cosmology, their ratios have been shown to have no dependence on the underlying cosmological model~\cite{Carter20}.
Therefore, considering these ratios, the model dependence on $r_d$ and $H_0$ disappears. 
We also obtain the error in the observed ratio mentioned above using error propagation. We consider the observables, with and without the estimated correlation coefficients. We then calculate the Bayes factors of the $\Lambda$CDM compared to the $R_h=ct$ model using {\tt Dynesty}~\cite{Speagle} software, which is based on the nested sampling algorithm. Since the $R_h=ct$ model has no free parameters, Bayesian evidence is the same as the likelihood. For the flat $\Lambda$CDM, we considered two sets of priors on $\Omega_M$: $\mathcal{U}$ (0,1) and $\mathcal{N}$(0.315,0.007), where the latter priors correspond to the best-fit parameters from Planck 2020 Cosmology~\cite{Planck2018}. A summary of the Bayes factors for both these priors for all the DESI DR1 observables can be found in Table~\ref{table1}. We find that for most observables up to a redshift of $z = 1.3$, the Bayes factors are close to 1, indicating that the significance of both models is comparable for each of these observables. However, when we consider the Lyman $\alpha$ QSO measurement at $z=2.33$, we find  Bayes factors ranging from 760 to 21500 for the four use cases, implying that $\Lambda$CDM is decisively favored compared to $R_h=ct$ using Jeffreys' scale~\cite{Trotta}. The same is the case when we do a combined analysis of all the DESI DR1 observables, where the Bayes factor provides decisive evidence for $\Lambda$CDM over $R_h=ct$. Our results for each of the observables do not change qualitatively, whether we include the correlation or not. They are also agnostic to the choice of the priors used.

We also show the ratio of the observed values of $D_m(z)/D_H(z)$ with and without correlations along with the expected theoretical curves for $\Lambda$CDM and $R_h=ct$. This plot can be found in Fig.~\ref{fig:ratioplot}. We can see that the Lyman-$\alpha$ measurement  at $z=2.33$ is consistent with $\Lambda$CDM and is  discrepant with respect to $R_h=ct$ by 4.5$\sigma$ (no correlation) and 4.1$\sigma$ (with correlation). The $\chi^2/dof$ values for the two plotted best fit $\Lambda$CDM models are 8.1/4 (with correlation) and 9.7/4(without correlations), corresponding to $p$-values of 0.09 and 0.05, respectively.
The corresponding numbers for  $R_h = ct$ are 32.3/5 (with correlations) and 26.5/5 (without correlations) with $p$-values of  $5.2 \times 10^{-6}$ and $7.1 \times 10^{-5}$, respectively. Therefore, although the flat $\Lambda$CDM does not provide a pristine fit to $D_M/D_H$, the $p$-values for $R_h=ct$ is about 3-5 orders of magnitude lower.  It is of course possible that extensions to $\Lambda$CDM such as wCDM or a time-varying dark energy equation of state provide a better fit to the ratio of $D_M$ by $D_H$ in DESI DRI. However, since our main goal is to compare the flat $\Lambda$CDM against $R_h=ct$, we do not pursue this further in this work.

Therefore, we find that the combined DESI DR1 BAO data conclusively rules our $R_h=ct$, although this is mainly driven by the Lyman $\alpha$ data point at $z=2.33$.
However, one conclusions are different compared to prveious BAO based tests my M23, which had  found that $R_h=ct$ is favored over $\Lambda$CDM.

\begin{table}[h!]
\begin{centering}

 \begin{tabular}{|c|c|c|c|c|c|}
    \hline
    Redshift & Tracers & Uniform and correlation & Uniform and no correlation & Normal and correlation & Normal and no correlation\\
    \hline
     0.51 & LRG1 & 0.7 & 0.8 & 0.1 & 0.1\\
     0.706 & LRG2 & 2.2 & 3.4 & 5.6 & 8.3\\
     0.93 & LRG3 + ELG1 & 0.6 & 0.7 & 2.9 & 3.6\\
     1.317 & ELG2 & 0.3 & 0.3 & 1.2 & 1.3\\
     2.33 & Ly$\alpha$ QSO & 762.7 & 4607.6 & 3226.1 & 21572.2\\
     Combined & & 788.5 & 5680.6 & 9699.5 & 77677.3\\
    \hline
 \end{tabular}
 \caption{\label{table1} Bayes Factor for $\Lambda$CDM cosmology wrt $R_h = ct$ cosmology. The terms ``Uniform'' and ``Normal'' imply Uniform and Gaussian prior on $\Omega_m$ for flat $\Lambda$CDM, where the Gaussian prior has been obtained from Planck cosmology~\cite{Planck2018} The sixth row shows the results for combined analysis of all redshifts. }
\end{centering}
\end{table}

\section{Conclusions}
About a year ago, the DESI Collaboration published its first data release of BAO measurements using galaxy, quasar, and Lyman-$\alpha$ forest tracers, spanning the redshift range from 0.1-4.2. We test  (along the same lines as M23) whether the ratio of $D_M$ to $D_H$ of DESI DR1 (normalized to the sound horizon) is compatible with $R_h=ct$ compared to   flat $\Lambda$CDM. For this purpose, we used Bayesian model comparison and evaluated the Bayes factor of  the flat $\Lambda$CDM compared to 
$R_h=ct$, since the latter model has no free parameters.
We use two sets of priors for flat $\Lambda$CDM and did the analysis with and without considering the correlations in the aforementioned ratio.

Our results of the Bayesian model comparison can be found in Table~\ref{table1}. We also show the measurements of the aforementioned ratio at different redshifts along with the theoretical expected curves for both models in Fig.~\ref{fig:ratioplot}. We find that when we separately consider each of the measurements until redshift of 1.3, both models are equally favored. However, when we consider the ratio at $z=2.33$ based on the Lyman-$\alpha$ measurement, we find that Bayes factors are higher than 100, indicating that  $\Lambda$CDM is decisively favored over $R_h=ct$ based on Jeffreys' scale. The same is the case when we combine all the measurements. \rthis{In case this measurement gets revised or if  systematics of this data point questioned, it would change our conclusion.}

Therefore, the DESI DR1 results for the ratio of transverse coming distance to Hubble distance are incompatible with $R_h=ct$, although this result is driven by the Lyman-$\alpha$ measurement at 
$z=2.33$. \rthis{In future works, we shall also test the robustness of this result by combing the DESI results with other geometric probes such as Type Ia Supernova and cosmic chronometers.}
\label{sec:conclusions}

\bibliography{References} 

\begin{thebibliography}{50}
\expandafter\ifx\csname natexlab\endcsname\relax\def\natexlab#1{#1}\fi
\expandafter\ifx\csname bibnamefont\endcsname\relax
  \def\bibnamefont#1{#1}\fi
\expandafter\ifx\csname bibfnamefont\endcsname\relax
  \def\bibfnamefont#1{#1}\fi
\expandafter\ifx\csname citenamefont\endcsname\relax
  \def\citenamefont#1{#1}\fi
\expandafter\ifx\csname url\endcsname\relax
  \def\url#1{\texttt{#1}}\fi
\expandafter\ifx\csname urlprefix\endcsname\relax\def\urlprefix{URL }\fi
\providecommand{\bibinfo}[2]{#2}
\providecommand{\eprint}[2][]{\url{#2}}

\bibitem[{\citenamefont{{Peebles} and {Ratra}}(2003)}]{Ratra03}
\bibinfo{author}{\bibfnamefont{P.~J.} \bibnamefont{{Peebles}}} \bibnamefont{and} \bibinfo{author}{\bibfnamefont{B.}~\bibnamefont{{Ratra}}}, \bibinfo{journal}{Reviews of Modern Physics} \textbf{\bibinfo{volume}{75}}, \bibinfo{pages}{559} (\bibinfo{year}{2003}), \eprint{astro-ph/0207347}.

\bibitem[{\citenamefont{{Planck Collaboration} et~al.}(2020)\citenamefont{{Planck Collaboration}, {Aghanim}, {Akrami}, {Ashdown}, {Aumont}, {Baccigalupi}, {Ballardini}, {Banday}, {Barreiro}, {Bartolo} et~al.}}]{Planck2018}
\bibinfo{author}{\bibnamefont{{Planck Collaboration}}}, \bibinfo{author}{\bibfnamefont{N.}~\bibnamefont{{Aghanim}}}, \bibinfo{author}{\bibfnamefont{Y.}~\bibnamefont{{Akrami}}}, \bibinfo{author}{\bibfnamefont{M.}~\bibnamefont{{Ashdown}}}, \bibinfo{author}{\bibfnamefont{J.}~\bibnamefont{{Aumont}}}, \bibinfo{author}{\bibfnamefont{C.}~\bibnamefont{{Baccigalupi}}}, \bibinfo{author}{\bibfnamefont{M.}~\bibnamefont{{Ballardini}}}, \bibinfo{author}{\bibfnamefont{A.~J.} \bibnamefont{{Banday}}}, \bibinfo{author}{\bibfnamefont{R.~B.} \bibnamefont{{Barreiro}}}, \bibinfo{author}{\bibfnamefont{N.}~\bibnamefont{{Bartolo}}}, \bibnamefont{et~al.}, \bibinfo{journal}{\aap} \textbf{\bibinfo{volume}{641}}, \bibinfo{eid}{A6} (\bibinfo{year}{2020}), \eprint{1807.06209}.

\bibitem[{\citenamefont{{Shah} et~al.}(2021)\citenamefont{{Shah}, {Lemos}, and {Lahav}}}]{Lahav}
\bibinfo{author}{\bibfnamefont{P.}~\bibnamefont{{Shah}}}, \bibinfo{author}{\bibfnamefont{P.}~\bibnamefont{{Lemos}}}, \bibnamefont{and} \bibinfo{author}{\bibfnamefont{O.}~\bibnamefont{{Lahav}}}, \bibinfo{journal}{\aapr} \textbf{\bibinfo{volume}{29}}, \bibinfo{eid}{9} (\bibinfo{year}{2021}), \eprint{2109.01161}.

\bibitem[{\citenamefont{{Di Valentino} et~al.}(2021)\citenamefont{{Di Valentino}, {Mena}, {Pan}, {Visinelli}, {Yang}, {Melchiorri}, {Mota}, {Riess}, and {Silk}}}]{DiValentino}
\bibinfo{author}{\bibfnamefont{E.}~\bibnamefont{{Di Valentino}}}, \bibinfo{author}{\bibfnamefont{O.}~\bibnamefont{{Mena}}}, \bibinfo{author}{\bibfnamefont{S.}~\bibnamefont{{Pan}}}, \bibinfo{author}{\bibfnamefont{L.}~\bibnamefont{{Visinelli}}}, \bibinfo{author}{\bibfnamefont{W.}~\bibnamefont{{Yang}}}, \bibinfo{author}{\bibfnamefont{A.}~\bibnamefont{{Melchiorri}}}, \bibinfo{author}{\bibfnamefont{D.~F.} \bibnamefont{{Mota}}}, \bibinfo{author}{\bibfnamefont{A.~G.} \bibnamefont{{Riess}}}, \bibnamefont{and} \bibinfo{author}{\bibfnamefont{J.}~\bibnamefont{{Silk}}}, \bibinfo{journal}{Classical and Quantum Gravity} \textbf{\bibinfo{volume}{38}}, \bibinfo{eid}{153001} (\bibinfo{year}{2021}), \eprint{2103.01183}.

\bibitem[{\citenamefont{{Verde} et~al.}(2019)\citenamefont{{Verde}, {Treu}, and {Riess}}}]{Verde}
\bibinfo{author}{\bibfnamefont{L.}~\bibnamefont{{Verde}}}, \bibinfo{author}{\bibfnamefont{T.}~\bibnamefont{{Treu}}}, \bibnamefont{and} \bibinfo{author}{\bibfnamefont{A.~G.} \bibnamefont{{Riess}}}, \bibinfo{journal}{Nature Astronomy} \textbf{\bibinfo{volume}{3}}, \bibinfo{pages}{891} (\bibinfo{year}{2019}), \eprint{1907.10625}.

\bibitem[{\citenamefont{Bethapudi and Desai}(2017)}]{Bethapudi}
\bibinfo{author}{\bibfnamefont{S.}~\bibnamefont{Bethapudi}} \bibnamefont{and} \bibinfo{author}{\bibfnamefont{S.}~\bibnamefont{Desai}}, \bibinfo{journal}{Eur. Phys. J. Plus} \textbf{\bibinfo{volume}{132}}, \bibinfo{pages}{78} (\bibinfo{year}{2017}), \eprint{1701.01789}.

\bibitem[{\citenamefont{{Cervantes-Cota} et~al.}(2023)\citenamefont{{Cervantes-Cota}, {Galindo-Uribarri}, and {Smoot}}}]{Smoot}
\bibinfo{author}{\bibfnamefont{J.~L.} \bibnamefont{{Cervantes-Cota}}}, \bibinfo{author}{\bibfnamefont{S.}~\bibnamefont{{Galindo-Uribarri}}}, \bibnamefont{and} \bibinfo{author}{\bibfnamefont{G.~F.} \bibnamefont{{Smoot}}}, \bibinfo{journal}{Universe} \textbf{\bibinfo{volume}{9}}, \bibinfo{eid}{501} (\bibinfo{year}{2023}), \eprint{2311.07552}.

\bibitem[{\citenamefont{{Abdalla} et~al.}(2022)\citenamefont{{Abdalla}, {Abell{\'a}n}, {Aboubrahim}, {Agnello}, {Akarsu}, {Akrami}, {Alestas}, {Aloni}, {Amendola}, {Anchordoqui} et~al.}}]{Abdalla22}
\bibinfo{author}{\bibfnamefont{E.}~\bibnamefont{{Abdalla}}}, \bibinfo{author}{\bibfnamefont{G.~F.} \bibnamefont{{Abell{\'a}n}}}, \bibinfo{author}{\bibfnamefont{A.}~\bibnamefont{{Aboubrahim}}}, \bibinfo{author}{\bibfnamefont{A.}~\bibnamefont{{Agnello}}}, \bibinfo{author}{\bibfnamefont{{\"O}.}~\bibnamefont{{Akarsu}}}, \bibinfo{author}{\bibfnamefont{Y.}~\bibnamefont{{Akrami}}}, \bibinfo{author}{\bibfnamefont{G.}~\bibnamefont{{Alestas}}}, \bibinfo{author}{\bibfnamefont{D.}~\bibnamefont{{Aloni}}}, \bibinfo{author}{\bibfnamefont{L.}~\bibnamefont{{Amendola}}}, \bibinfo{author}{\bibfnamefont{L.~A.} \bibnamefont{{Anchordoqui}}}, \bibnamefont{et~al.}, \bibinfo{journal}{Journal of High Energy Astrophysics} \textbf{\bibinfo{volume}{34}}, \bibinfo{pages}{49} (\bibinfo{year}{2022}), \eprint{2203.06142}.

\bibitem[{\citenamefont{{Merritt}}(2017)}]{Merritt}
\bibinfo{author}{\bibfnamefont{D.}~\bibnamefont{{Merritt}}}, \bibinfo{journal}{Studies in the History and Philosophy of Modern Physics} \textbf{\bibinfo{volume}{57}}, \bibinfo{pages}{41} (\bibinfo{year}{2017}), \eprint{1703.02389}.

\bibitem[{\citenamefont{{Fields} et~al.}(2020)\citenamefont{{Fields}, {Olive}, {Yeh}, and {Young}}}]{Fields}
\bibinfo{author}{\bibfnamefont{B.~D.} \bibnamefont{{Fields}}}, \bibinfo{author}{\bibfnamefont{K.~A.} \bibnamefont{{Olive}}}, \bibinfo{author}{\bibfnamefont{T.-H.} \bibnamefont{{Yeh}}}, \bibnamefont{and} \bibinfo{author}{\bibfnamefont{C.}~\bibnamefont{{Young}}}, \bibinfo{journal}{\jcap} \textbf{\bibinfo{volume}{2020}}, \bibinfo{eid}{010} (\bibinfo{year}{2020}), \eprint{1912.01132}.

\bibitem[{\citenamefont{{Boylan-Kolchin}}(2023)}]{Boylan}
\bibinfo{author}{\bibfnamefont{M.}~\bibnamefont{{Boylan-Kolchin}}}, \bibinfo{journal}{Nature Astronomy} \textbf{\bibinfo{volume}{7}}, \bibinfo{pages}{731} (\bibinfo{year}{2023}), \eprint{2208.01611}.

\bibitem[{\citenamefont{{Perivolaropoulos} and {Skara}}(2022)}]{Periv}
\bibinfo{author}{\bibfnamefont{L.}~\bibnamefont{{Perivolaropoulos}}} \bibnamefont{and} \bibinfo{author}{\bibfnamefont{F.}~\bibnamefont{{Skara}}}, \bibinfo{journal}{\nar} \textbf{\bibinfo{volume}{95}}, \bibinfo{eid}{101659} (\bibinfo{year}{2022}), \eprint{2105.05208}.

\bibitem[{\citenamefont{{Peebles}}(2022)}]{Peebles22}
\bibinfo{author}{\bibfnamefont{P.~J.~E.} \bibnamefont{{Peebles}}}, \bibinfo{journal}{Annals of Physics} \textbf{\bibinfo{volume}{447}}, \bibinfo{eid}{169159} (\bibinfo{year}{2022}), \eprint{2208.05018}.

\bibitem[{\citenamefont{{Banik} and {Zhao}}(2022)}]{Banik}
\bibinfo{author}{\bibfnamefont{I.}~\bibnamefont{{Banik}}} \bibnamefont{and} \bibinfo{author}{\bibfnamefont{H.}~\bibnamefont{{Zhao}}}, \bibinfo{journal}{Symmetry} \textbf{\bibinfo{volume}{14}}, \bibinfo{eid}{1331} (\bibinfo{year}{2022}), \eprint{2110.06936}.

\bibitem[{\citenamefont{{Melia}}(2007)}]{Melia07}
\bibinfo{author}{\bibfnamefont{F.}~\bibnamefont{{Melia}}}, \bibinfo{journal}{\mnras} \textbf{\bibinfo{volume}{382}}, \bibinfo{pages}{1917} (\bibinfo{year}{2007}), \eprint{0711.4181}.

\bibitem[{\citenamefont{Melia and Shevchuk}(2012)}]{Shevchuk}
\bibinfo{author}{\bibfnamefont{F.}~\bibnamefont{Melia}} \bibnamefont{and} \bibinfo{author}{\bibfnamefont{A.}~\bibnamefont{Shevchuk}}, \bibinfo{journal}{Mon. Not. Roy. Astron. Soc.} \textbf{\bibinfo{volume}{419}}, \bibinfo{pages}{2579} (\bibinfo{year}{2012}), \eprint{1109.5189}.

\bibitem[{\citenamefont{{Melia}}(2012)}]{Melia2012}
\bibinfo{author}{\bibfnamefont{F.}~\bibnamefont{{Melia}}}, \bibinfo{journal}{arXiv e-prints} \bibinfo{eid}{arXiv:1205.2713} (\bibinfo{year}{2012}), \eprint{1205.2713}.

\bibitem[{\citenamefont{{Melia} and {Maier}}(2013)}]{Meliamaier}
\bibinfo{author}{\bibfnamefont{F.}~\bibnamefont{{Melia}}} \bibnamefont{and} \bibinfo{author}{\bibfnamefont{R.~S.} \bibnamefont{{Maier}}}, \bibinfo{journal}{\mnras} \textbf{\bibinfo{volume}{432}}, \bibinfo{pages}{2669} (\bibinfo{year}{2013}), \eprint{1304.1802}.

\bibitem[{\citenamefont{{Melia} and {Yennapureddy}}(2018)}]{Meliachrono}
\bibinfo{author}{\bibfnamefont{F.}~\bibnamefont{{Melia}}} \bibnamefont{and} \bibinfo{author}{\bibfnamefont{M.~K.} \bibnamefont{{Yennapureddy}}}, \bibinfo{journal}{\jcap} \textbf{\bibinfo{volume}{2018}}, \bibinfo{eid}{034} (\bibinfo{year}{2018}), \eprint{1802.02255}.

\bibitem[{\citenamefont{{Wan} et~al.}(2019)\citenamefont{{Wan}, {Cao}, {Melia}, and {Zhang}}}]{Meliaquasar}
\bibinfo{author}{\bibfnamefont{H.-Y.} \bibnamefont{{Wan}}}, \bibinfo{author}{\bibfnamefont{S.-L.} \bibnamefont{{Cao}}}, \bibinfo{author}{\bibfnamefont{F.}~\bibnamefont{{Melia}}}, \bibnamefont{and} \bibinfo{author}{\bibfnamefont{T.-J.} \bibnamefont{{Zhang}}}, \bibinfo{journal}{Physics of the Dark Universe} \textbf{\bibinfo{volume}{26}}, \bibinfo{eid}{100405} (\bibinfo{year}{2019}), \eprint{1910.14024}.

\bibitem[{\citenamefont{{Melia}}(2019)}]{MeliaUV}
\bibinfo{author}{\bibfnamefont{F.}~\bibnamefont{{Melia}}}, \bibinfo{journal}{\mnras} \textbf{\bibinfo{volume}{489}}, \bibinfo{pages}{517} (\bibinfo{year}{2019}), \eprint{1907.13127}.

\bibitem[{\citenamefont{{Melia}}(2016)}]{Meliafgas}
\bibinfo{author}{\bibfnamefont{F.}~\bibnamefont{{Melia}}}, \bibinfo{journal}{Proceedings of the Royal Society of London Series A} \textbf{\bibinfo{volume}{472}}, \bibinfo{pages}{20150765} (\bibinfo{year}{2016}), \eprint{1601.04649}.

\bibitem[{\citenamefont{{Melia} et~al.}(2018)\citenamefont{{Melia}, {Wei}, {Maier}, and {Wu}}}]{MeliaSN}
\bibinfo{author}{\bibfnamefont{F.}~\bibnamefont{{Melia}}}, \bibinfo{author}{\bibfnamefont{J.~J.} \bibnamefont{{Wei}}}, \bibinfo{author}{\bibfnamefont{R.~S.} \bibnamefont{{Maier}}}, \bibnamefont{and} \bibinfo{author}{\bibfnamefont{X.~F.} \bibnamefont{{Wu}}}, \bibinfo{journal}{EPL (Europhysics Letters)} \textbf{\bibinfo{volume}{123}}, \bibinfo{pages}{59002} (\bibinfo{year}{2018}), \eprint{1809.05094}.

\bibitem[{\citenamefont{{Leaf} and {Melia}}(2018)}]{MeliaSL}
\bibinfo{author}{\bibfnamefont{K.}~\bibnamefont{{Leaf}}} \bibnamefont{and} \bibinfo{author}{\bibfnamefont{F.}~\bibnamefont{{Melia}}}, \bibinfo{journal}{\mnras} \textbf{\bibinfo{volume}{478}}, \bibinfo{pages}{5104} (\bibinfo{year}{2018}), \eprint{1805.08640}.

\bibitem[{\citenamefont{{Melia} et~al.}(2023)\citenamefont{{Melia}, {Wei}, and {Wu}}}]{Melialensing23}
\bibinfo{author}{\bibfnamefont{F.}~\bibnamefont{{Melia}}}, \bibinfo{author}{\bibfnamefont{J.-J.} \bibnamefont{{Wei}}}, \bibnamefont{and} \bibinfo{author}{\bibfnamefont{X.-F.} \bibnamefont{{Wu}}}, \bibinfo{journal}{\mnras} \textbf{\bibinfo{volume}{519}}, \bibinfo{pages}{2528} (\bibinfo{year}{2023}), \eprint{2212.06113}.

\bibitem[{\citenamefont{{Wei} and {Melia}}(2023)}]{MeliaFRB}
\bibinfo{author}{\bibfnamefont{J.-J.} \bibnamefont{{Wei}}} \bibnamefont{and} \bibinfo{author}{\bibfnamefont{F.}~\bibnamefont{{Melia}}}, \bibinfo{journal}{\apj} \textbf{\bibinfo{volume}{955}}, \bibinfo{eid}{101} (\bibinfo{year}{2023}), \eprint{2308.05918}.

\bibitem[{\citenamefont{{Melia}}(2024)}]{MeliaJWST}
\bibinfo{author}{\bibfnamefont{F.}~\bibnamefont{{Melia}}}, \bibinfo{journal}{European Physical Journal C} \textbf{\bibinfo{volume}{84}}, \bibinfo{eid}{1279} (\bibinfo{year}{2024}), \eprint{2412.02706}.

\bibitem[{\citenamefont{{Melia}}(2025)}]{MeliaJWST1}
\bibinfo{author}{\bibfnamefont{F.}~\bibnamefont{{Melia}}}, \bibinfo{journal}{arXiv e-prints} \bibinfo{eid}{arXiv:2503.09607} (\bibinfo{year}{2025}), \eprint{2503.09607}.

\bibitem[{\citenamefont{{Shafer}}(2015)}]{Shafer}
\bibinfo{author}{\bibfnamefont{D.~L.} \bibnamefont{{Shafer}}}, \bibinfo{journal}{\prd} \textbf{\bibinfo{volume}{91}}, \bibinfo{eid}{103516} (\bibinfo{year}{2015}), \eprint{1502.05416}.

\bibitem[{\citenamefont{{Bilicki} and {Seikel}}(2012)}]{Seikel_2012_rhct}
\bibinfo{author}{\bibfnamefont{M.}~\bibnamefont{{Bilicki}}} \bibnamefont{and} \bibinfo{author}{\bibfnamefont{M.}~\bibnamefont{{Seikel}}}, \bibinfo{journal}{\mnras} \textbf{\bibinfo{volume}{425}}, \bibinfo{pages}{1664} (\bibinfo{year}{2012}), \eprint{1206.5130}.

\bibitem[{\citenamefont{{Lewis} et~al.}(2016)\citenamefont{{Lewis}, {Barnes}, and {Kaushik}}}]{LewisBBN}
\bibinfo{author}{\bibfnamefont{G.~F.} \bibnamefont{{Lewis}}}, \bibinfo{author}{\bibfnamefont{L.~A.} \bibnamefont{{Barnes}}}, \bibnamefont{and} \bibinfo{author}{\bibfnamefont{R.}~\bibnamefont{{Kaushik}}}, \bibinfo{journal}{\mnras} \textbf{\bibinfo{volume}{460}}, \bibinfo{pages}{291} (\bibinfo{year}{2016}), \eprint{1604.07460}.

\bibitem[{\citenamefont{{Haridasu} et~al.}(2017)\citenamefont{{Haridasu}, {Lukovi{\'c}}, {D'Agostino}, and {Vittorio}}}]{Haridasu1}
\bibinfo{author}{\bibfnamefont{B.~S.} \bibnamefont{{Haridasu}}}, \bibinfo{author}{\bibfnamefont{V.~V.} \bibnamefont{{Lukovi{\'c}}}}, \bibinfo{author}{\bibfnamefont{R.}~\bibnamefont{{D'Agostino}}}, \bibnamefont{and} \bibinfo{author}{\bibfnamefont{N.}~\bibnamefont{{Vittorio}}}, \bibinfo{journal}{\aap} \textbf{\bibinfo{volume}{600}}, \bibinfo{eid}{L1} (\bibinfo{year}{2017}), \eprint{1702.08244}.

\bibitem[{\citenamefont{{Lin} et~al.}(2018)\citenamefont{{Lin}, {Li}, and {Sang}}}]{Lin}
\bibinfo{author}{\bibfnamefont{H.-N.} \bibnamefont{{Lin}}}, \bibinfo{author}{\bibfnamefont{X.}~\bibnamefont{{Li}}}, \bibnamefont{and} \bibinfo{author}{\bibfnamefont{Y.}~\bibnamefont{{Sang}}}, \bibinfo{journal}{Chinese Physics C} \textbf{\bibinfo{volume}{42}}, \bibinfo{eid}{095101} (\bibinfo{year}{2018}), \eprint{1711.05025}.

\bibitem[{\citenamefont{{Hu} and {Wang}}(2018)}]{Hu2018}
\bibinfo{author}{\bibfnamefont{J.}~\bibnamefont{{Hu}}} \bibnamefont{and} \bibinfo{author}{\bibfnamefont{F.~Y.} \bibnamefont{{Wang}}}, \bibinfo{journal}{\mnras} \textbf{\bibinfo{volume}{477}}, \bibinfo{pages}{5064} (\bibinfo{year}{2018}), \eprint{1804.06606}.

\bibitem[{\citenamefont{{Tu} et~al.}(2019)\citenamefont{{Tu}, {Hu}, and {Wang}}}]{Tu2019}
\bibinfo{author}{\bibfnamefont{Z.~L.} \bibnamefont{{Tu}}}, \bibinfo{author}{\bibfnamefont{J.}~\bibnamefont{{Hu}}}, \bibnamefont{and} \bibinfo{author}{\bibfnamefont{F.~Y.} \bibnamefont{{Wang}}}, \bibinfo{journal}{\mnras} \textbf{\bibinfo{volume}{484}}, \bibinfo{pages}{4337} (\bibinfo{year}{2019}), \eprint{1901.09144}.

\bibitem[{\citenamefont{{Fujii}}(2020)}]{Fuji}
\bibinfo{author}{\bibfnamefont{H.}~\bibnamefont{{Fujii}}}, \bibinfo{journal}{Research Notes of the American Astronomical Society} \textbf{\bibinfo{volume}{4}}, \bibinfo{eid}{72} (\bibinfo{year}{2020}).

\bibitem[{\citenamefont{{Singirikonda} and {Desai}}(2020)}]{Haveesh}
\bibinfo{author}{\bibfnamefont{H.}~\bibnamefont{{Singirikonda}}} \bibnamefont{and} \bibinfo{author}{\bibfnamefont{S.}~\bibnamefont{{Desai}}}, \bibinfo{journal}{European Physical Journal C} \textbf{\bibinfo{volume}{80}}, \bibinfo{eid}{694} (\bibinfo{year}{2020}), \eprint{2003.00494}.

\bibitem[{\citenamefont{{Panchal} and {Desai}}(2024)}]{Panchalfgas}
\bibinfo{author}{\bibfnamefont{K.}~\bibnamefont{{Panchal}}} \bibnamefont{and} \bibinfo{author}{\bibfnamefont{S.}~\bibnamefont{{Desai}}}, \bibinfo{journal}{Journal of High Energy Astrophysics} \textbf{\bibinfo{volume}{43}}, \bibinfo{pages}{15} (\bibinfo{year}{2024}), \eprint{2401.11138}.

\bibitem[{\citenamefont{{Melia} and {L{\'o}pez-Corredoira}}(2017)}]{MeliaBAO17}
\bibinfo{author}{\bibfnamefont{F.}~\bibnamefont{{Melia}}} \bibnamefont{and} \bibinfo{author}{\bibfnamefont{M.}~\bibnamefont{{L{\'o}pez-Corredoira}}}, \bibinfo{journal}{International Journal of Modern Physics D} \textbf{\bibinfo{volume}{26}}, \bibinfo{eid}{1750055-265} (\bibinfo{year}{2017}), \eprint{1503.05052}.

\bibitem[{\citenamefont{{Melia} et~al.}(2020)\citenamefont{{Melia}, {Qin}, and {Zhang}}}]{MeliaBAO20}
\bibinfo{author}{\bibfnamefont{F.}~\bibnamefont{{Melia}}}, \bibinfo{author}{\bibfnamefont{J.}~\bibnamefont{{Qin}}}, \bibnamefont{and} \bibinfo{author}{\bibfnamefont{T.-J.} \bibnamefont{{Zhang}}}, \bibinfo{journal}{\mnras} \textbf{\bibinfo{volume}{499}}, \bibinfo{pages}{L36} (\bibinfo{year}{2020}), \eprint{2008.12628}.

\bibitem[{\citenamefont{{Melia} and {L{\'o}pez-Corredoira}}(2022)}]{MeliaBAO22}
\bibinfo{author}{\bibfnamefont{F.}~\bibnamefont{{Melia}}} \bibnamefont{and} \bibinfo{author}{\bibfnamefont{M.}~\bibnamefont{{L{\'o}pez-Corredoira}}}, \bibinfo{journal}{International Journal of Modern Physics D} \textbf{\bibinfo{volume}{31}}, \bibinfo{eid}{2250065} (\bibinfo{year}{2022}), \eprint{2204.02186}.

\bibitem[{\citenamefont{{Melia}}(2023)}]{Melia23}
\bibinfo{author}{\bibfnamefont{F.}~\bibnamefont{{Melia}}}, \bibinfo{journal}{EPL (Europhysics Letters)} \textbf{\bibinfo{volume}{143}}, \bibinfo{eid}{59004} (\bibinfo{year}{2023}), \eprint{2309.00662}.

\bibitem[{\citenamefont{{DESI Collaboration} et~al.}(2025)\citenamefont{{DESI Collaboration}, {Abdul-Karim}, {Aguilar}, {Ahlen}, {Alam}, {Allen}, {Allende Prieto}, {Alves}, {Anand}, {Andrade} et~al.}}]{DESI2025}
\bibinfo{author}{\bibnamefont{{DESI Collaboration}}}, \bibinfo{author}{\bibfnamefont{M.}~\bibnamefont{{Abdul-Karim}}}, \bibinfo{author}{\bibfnamefont{J.}~\bibnamefont{{Aguilar}}}, \bibinfo{author}{\bibfnamefont{S.}~\bibnamefont{{Ahlen}}}, \bibinfo{author}{\bibfnamefont{S.}~\bibnamefont{{Alam}}}, \bibinfo{author}{\bibfnamefont{L.}~\bibnamefont{{Allen}}}, \bibinfo{author}{\bibfnamefont{C.}~\bibnamefont{{Allende Prieto}}}, \bibinfo{author}{\bibfnamefont{O.}~\bibnamefont{{Alves}}}, \bibinfo{author}{\bibfnamefont{A.}~\bibnamefont{{Anand}}}, \bibinfo{author}{\bibfnamefont{U.}~\bibnamefont{{Andrade}}}, \bibnamefont{et~al.}, \bibinfo{journal}{arXiv e-prints} \bibinfo{eid}{arXiv:2503.14738} (\bibinfo{year}{2025}), \eprint{2503.14738}.

\bibitem[{\citenamefont{{Adame} et~al.}(2025)\citenamefont{{Adame}, {Aguilar}, {Ahlen}, {Alam}, {Alexander}, {Alvarez}, {Alves}, {Anand}, {Andrade}, {Armengaud} et~al.}}]{DESI2024}
\bibinfo{author}{\bibfnamefont{A.~G.} \bibnamefont{{Adame}}}, \bibinfo{author}{\bibfnamefont{J.}~\bibnamefont{{Aguilar}}}, \bibinfo{author}{\bibfnamefont{S.}~\bibnamefont{{Ahlen}}}, \bibinfo{author}{\bibfnamefont{S.}~\bibnamefont{{Alam}}}, \bibinfo{author}{\bibfnamefont{D.~M.} \bibnamefont{{Alexander}}}, \bibinfo{author}{\bibfnamefont{M.}~\bibnamefont{{Alvarez}}}, \bibinfo{author}{\bibfnamefont{O.}~\bibnamefont{{Alves}}}, \bibinfo{author}{\bibfnamefont{A.}~\bibnamefont{{Anand}}}, \bibinfo{author}{\bibfnamefont{U.}~\bibnamefont{{Andrade}}}, \bibinfo{author}{\bibfnamefont{E.}~\bibnamefont{{Armengaud}}}, \bibnamefont{et~al.}, \bibinfo{journal}{\jcap} \textbf{\bibinfo{volume}{2025}}, \bibinfo{eid}{021} (\bibinfo{year}{2025}), \eprint{2404.03002}.

\bibitem[{\citenamefont{{Trotta}}(2017)}]{Trotta}
\bibinfo{author}{\bibfnamefont{R.}~\bibnamefont{{Trotta}}}, \bibinfo{journal}{ArXiv e-prints}  (\bibinfo{year}{2017}), \eprint{1701.01467}.

\bibitem[{\citenamefont{{Kerscher} and {Weller}}(2019)}]{Weller}
\bibinfo{author}{\bibfnamefont{M.}~\bibnamefont{{Kerscher}}} \bibnamefont{and} \bibinfo{author}{\bibfnamefont{J.}~\bibnamefont{{Weller}}}, \bibinfo{journal}{SciPost Physics Lecture Notes} \textbf{\bibinfo{volume}{9}} (\bibinfo{year}{2019}), \eprint{1901.07726}.

\bibitem[{\citenamefont{{Sharma}}(2017)}]{Sanjib}
\bibinfo{author}{\bibfnamefont{S.}~\bibnamefont{{Sharma}}}, \bibinfo{journal}{\araa} \textbf{\bibinfo{volume}{55}}, \bibinfo{pages}{213} (\bibinfo{year}{2017}), \eprint{1706.01629}.

\bibitem[{\citenamefont{{Krishak} and {Desai}}(2020)}]{Krishak}
\bibinfo{author}{\bibfnamefont{A.}~\bibnamefont{{Krishak}}} \bibnamefont{and} \bibinfo{author}{\bibfnamefont{S.}~\bibnamefont{{Desai}}}, \bibinfo{journal}{\jcap} \textbf{\bibinfo{volume}{2020}}, \bibinfo{eid}{006} (\bibinfo{year}{2020}), \eprint{2003.10127}.

\bibitem[{\citenamefont{{Carter} et~al.}(2020)\citenamefont{{Carter}, {Beutler}, {Percival}, {DeRose}, {Wechsler}, and {Zhao}}}]{Carter20}
\bibinfo{author}{\bibfnamefont{P.}~\bibnamefont{{Carter}}}, \bibinfo{author}{\bibfnamefont{F.}~\bibnamefont{{Beutler}}}, \bibinfo{author}{\bibfnamefont{W.~J.} \bibnamefont{{Percival}}}, \bibinfo{author}{\bibfnamefont{J.}~\bibnamefont{{DeRose}}}, \bibinfo{author}{\bibfnamefont{R.~H.} \bibnamefont{{Wechsler}}}, \bibnamefont{and} \bibinfo{author}{\bibfnamefont{C.}~\bibnamefont{{Zhao}}}, \bibinfo{journal}{\mnras} \textbf{\bibinfo{volume}{494}}, \bibinfo{pages}{2076} (\bibinfo{year}{2020}), \eprint{1906.03035}.

\bibitem[{\citenamefont{{Speagle}}(2020)}]{Speagle}
\bibinfo{author}{\bibfnamefont{J.~S.} \bibnamefont{{Speagle}}}, \bibinfo{journal}{\mnras} \textbf{\bibinfo{volume}{493}}, \bibinfo{pages}{3132} (\bibinfo{year}{2020}), \eprint{1904.02180}.

\end{thebibliography}
\end{document}